\documentclass[12pt]{article}

\newcommand{\bbr}{I\!\! R}

\newcommand{\x}{arXiv:}
\newcommand{\m}{\mathrm}

\usepackage{graphicx}

\begin{document}
\thispagestyle{empty}
\begin{center}

\null \vskip-1truecm \vskip2truecm

{\Large{\bf \textsf{Horizon Complementarity and Casimir Violations
of the }}}

\bigskip

{\Large{\bf \textsf{Null Energy Condition}}}

{\large{\bf \textsf{}}}

\vskip1truecm

{\large \textsf{Brett McInnes}}

\vskip1truecm

\textsf{\\  National
  University of Singapore}

\textsf{email: matmcinn@nus.edu.sg}\\

\end{center}
\vskip1truecm \centerline{\textsf{ABSTRACT}} \baselineskip=15pt
\medskip

The principle of \emph{horizon complementarity} is an attempt to
extend ideas about black hole complementarity to all horizons,
including cosmological ones. The idea is that the degrees of freedom
necessary to describe the interior of the cosmic horizon of
\emph{one} observer in a given universe are in fact sufficient to
account for the physics of that entire universe: the remainder is
just a set of redundant copies of the interior of a single cosmic
horizon. \emph{These copies must be factored out}, just as one has
to factor out gauge redundancies to identify the true degrees of
freedom in gauge theory. Motivated by the observation that quantum
cosmology favours \emph{compactified} negatively curved spatial
sections, we propose to use such geometries to implement horizon
complementarity for eternal Inflation. We point out that the
``effective finiteness" of such universes has important consequences
for physics \emph{inside} the observer's horizon: there is a
non-local effect, represented by a Casimir energy. We use our
proposed interpretation of complementarity to constrain the
gravitational Casimir coupling in two very different ways; the
result is an explicit prediction for the value of the coupling.

\newpage

\addtocounter{section}{1}
\section* {\large{\textsf{1. Horizon Complementarity Applied to Cosmology}}}

The most fully developed proposed explanation of the value of the
cosmological constant \cite{kn:polchinski}\cite{kn:TASI} is based on
the Coleman-De Luccia bubble nucleation mechanism
\cite{kn:deluccia}\cite{kn:frei}. Each nucleation can reduce the
value of the vacuum energy; it turns out that the range of possible
values is a ``discretuum" with gaps small enough to account for the
value observed in our Universe\footnote{The basic reference is
\cite{kn:boupolch}; see for example
\cite{kn:tye}\cite{kn:podolsky1}\cite{kn:wrinkle}\cite{kn:sash} for
a selection of more recent work on this theme.}.

Once we have a concrete mechanism for producing a Universe which
resembles the one we observe, we should be able to use it to work
towards answers to some fundamental questions. In particular, we
should ultimately obtain a specific picture of the structure of our
Universe in its earliest stages, even before Inflation
\cite{kn:baby}.

The standard picture of this part of cosmic history is as follows.
The passage through the bubble wall mediates a change in the set of
distinguished cosmological observers, and the internal observers see
a classical spacetime which is infinite to the future [if the vacuum
energy is still positive ---$\,$ it is then future asymptotically de
Sitter] and also \cite{kn:frei} in space: the spatial sections are
copies of three-dimensional \emph{hyperbolic space}, H$^3$.

It has long been known that infinite space gives rise to various
apparent paradoxes, which continue to elude a generally accepted
solution\footnote{For clear discussions and lists of references
see \cite{kn:pagenew}\cite{kn:sheng}\cite{kn:simone}.}. It is
natural to ask: is the standard interpretation of bubble spacetime
geometry misleading us? Could it be that bubble worlds are, in
some sense, \emph{effectively finite}
\cite{kn:pagedecaying}\cite{kn:freebird} when non-classical
effects are taken into account? If so, the most serious
difficulties of eternal Inflation would be resolved at a stroke.
We shall argue here that, in particular, the question of
\emph{spatial} finiteness is relevant, indeed essential, to a full
account of the history of the \emph{observable} Universe, whether
or not the finiteness is itself in any way directly
observable\footnote{Throughout this work, we assume that the
[spatial] finiteness of the Universe is \emph{not} directly
observable at the present time \cite{kn:janna}, or indeed at any
time. This is natural if one uses Inflation, as we do here; see
however \cite{kn:tavakol} for a recent discussion of relevant
observations.}.

As is well known, in an asymptotically de Sitter spacetime no single
observer can ever receive signals from more than a finite region of
any spatial section: there are cosmological horizons. The spatial
sections are only infinite in a physical sense if one takes a
\emph{global} perspective, that is, if one imagines collating the
observations of an infinite family of observers. As is well known,
taking such a global point of view leads to serious difficulties in
the case of black hole horizons \cite{kn:comp}, and it is natural to
suggest
\cite{kn:dyson}\cite{kn:savon}\cite{kn:trouble}\cite{kn:bouff} that
analogous problems might arise if one tries to do so here.

In this way one is led to the concept of \emph{horizon
complementarity}. This is the natural generalization of the
well-known principle of black hole complementarity \cite{kn:comp} to
\emph{all} horizons: one should not expect to be able to give a
consistent account of the physics on both sides of \emph{any}
horizon. In other words, the degrees of freedom used by \emph{one}
observer are sufficient to describe an entire universe, of whatever
size. This is the sense in which an infinite world can be
``effectively finite".

The simplicity and reasonableness of this proposal, which just
excludes all data that can never affect a given observer, are
evident\footnote{Note that objects outside the bubble can also have
a direct effect on conditions \emph{inside} the cosmological
horizon: see \cite{kn:fish}\cite{kn:klebby}\cite{kn:wrath}. This has
no bearing on the question being discussed here.}. It is,
nevertheless, a very radical departure: the claim is that an
infinite universe can be described by the degrees of freedom of a
finite one. We are entitled to ask how this works in more detail.
One would also like to know whether this drastic reduction has any
consequences [through some non-local effect] for the earliest
history of that part of the Universe which \emph{can} affect the
observer.

In this work, we shall propose an approach to answering such
questions. The basic idea is that the effective spatial finiteness
of the horizon interior can be expressed concretely in terms of the
way the hyperbolic space H$^3$ can be reduced to a compact quotient
\cite{kn:thurston} of the form H$^3$/$\Gamma$, where $\Gamma$ is one
of the many infinite discrete subgroups of O(1,3) that can act
freely, isometrically, and properly discontinuously on H$^3$. This
``taking of a quotient" amounts to declaring that the infinite bulk
of H$^3$ represents an infinitely \emph{redundant description} of a
particular finite sub-domain $\Delta$, a fundamental domain
containing an observer and defined by $\Gamma$. To understand this,
note that the faces of the fundamental domain are topologically
identified in pairs, so that an observer venturing beyond the
boundary of $\Delta$ can never find anything ``new". From the
perspective of the fundamental domain, the full hyperbolic space
consists of an infinite number of identical copies of $\Delta$,
joined along their boundaries in a manner described by the discrete
rotations contained in $\Gamma$.

We therefore claim that $\Gamma$ is like a group of \emph{gauge
transformations}, which, like any other such group, must be factored
out in order to identify the true physical degrees of
freedom\footnote{The idea that the spatial sections of bubble
universes might not be globally identical to H$^3$ was suggested in
\cite{kn:hsu1}; the modification suggested there is however very
much more drastic than what we are proposing here.}. When the
process is described in this way, it is clear that this is a precise
mathematical implementation of the concepts underlying horizon
complementarity, as it applies to Coleman-De Luccia bubble
universes. Note that spatial sections of the form H$^3$/$\Gamma$ are
in fact very natural from a string-theoretic point of view: string
winding ``transforms geometry to topology" exactly when one
compactifies H$^3$ in the way we propose here: see
\cite{kn:starr}\cite{kn:silvery}.

This interpretation of horizon complementarity is suggested by
recent discussions of the quantum creation of a universe from
``nothing". It appears that quantum-gravitational processes in this
case do favour compact spatial sections, and in fact they favour
compact \emph{negatively curved} [or perhaps flat] sections. This
has been argued with various emphases by Zeldovich and Starobinsky
\cite{kn:starobinsky}, by Coule and Martin \cite{kn:martin}, and by
Linde \cite{kn:lindetypical}\cite{kn:lindenew}. Linde in particular
stresses that in the case of spacetimes produced by quantum creation
from ``nothing", compact negatively curved spatial sections are
likely to be favoured, because in this case there is no barrier
through which one must tunnel, and because \emph{finite worlds are
easier to create}. We are essentially proposing to adapt some of the
lessons learnt in these studies to the Coleman-De Luccia case.

The reduction of H$^3$ to H$^3/\Gamma$ gives us a concrete
mathematical basis for investigating the physical consequences of
horizon complementarity, as applied to the \emph{inflationary}
horizons established immediately after the bubble spacetime emerges
from the bubble wall. In fact, we will argue that this
implementation of horizon complementarity leads to an extremely
specific description of the geometry of the bubble spacetime in its
earliest stages. The argument runs as follows. As is well known, the
compactification of spatial sections leads to a cosmological version
of the Casimir effect. In a Friedmann cosmological model, the
Casimir energy is represented by a term in which the Casimir
coupling to gravity is described by a certain constant, denoted by
$\alpha$ here. We show that the requirement of non-perturbative
stability [in string theory] puts an upper bound on $\alpha$, while
horizon complementarity itself puts a direct lower bound on it. The
range of $\alpha$ values compatible with both constraints proves to
be extremely narrow, thus effectively allowing us to compute
$\alpha$. In this way we can specify the spacetime metric of the
earliest [bubble] spacetime quite precisely.

We begin by describing our proposal in greater detail.

\addtocounter{section}{1}
\section* {\large{\textsf{2. The Proposal in More Detail}}}
Let us begin by examining the conformal diagram representing bubble
nucleation [Figure 1].
\begin{figure}[!h]
\centering
\includegraphics[width=0.8\textwidth]{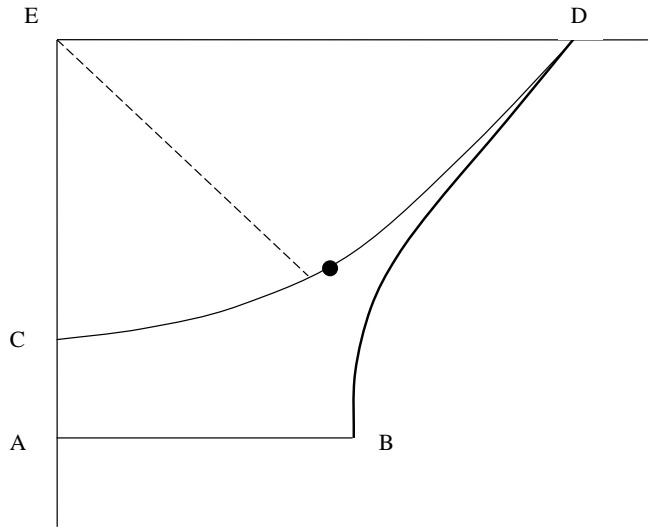}
\caption{Thick-Walled Bubble Universe.}
\end{figure}
Note that bubble nucleation cannot be represented in an entirely
classical manner, but the essential points are still captured by the
diagram. Nucleation occurs along AB, and the spacetime trajectory of
the outer surface of the bubble wall is represented by BD. Following
Aguirre and Gratton \cite{kn:aguirregratton}, we think of bubble
nucleation as a process which separates the spacetime into three
zones: the original spacetime and the bubble interior [ECD in the
diagram] can be described more or less accurately in a
semi-classical way, but the interior of the bubble wall [ACDB in the
diagram] is a predominantly quantum domain. As with any quantum
process [like the decay of an unstable particle, as discussed by
Aguirre and Gratton], the region ACDB can be regarded as a system
which \emph{prepares the initial conditions for the subsequent
semi-classical evolution}. We do not know how to describe the
spacetime geometry inside ACDB; but the spacelike surface CD must
still bear the imprint of the non-classical physics [such as that
which gives rise to horizon complementarity] which govern the
interior of the bubble wall. The semi-classical spacetime inside the
bubble has to be \emph{cut off} along CD; but this will not be done
in an arbitrary way: it will be done in a manner compatible with
whatever we know about the properties of the strictly quantum domain
ACDB.

We propose that horizon complementarity can be formulated in the
following way: we think of CD as being broken up into an infinite
collection of identical copies of a fixed finite \emph{fundamental
domain} $\Delta$ \cite{kn:thurston}, and declare that the physical
degrees of freedom associated with $\Delta$ suffice to describe CD
completely. If we think of $\Delta$ as being centred on the origin
of coordinates, its boundary is symbolised by the heavy dot on CD in
Figure 1. [Other possible locations for $\Delta$ can be pictured as
intervals on CD, which would appear to become smaller and smaller as
D is approached. Note that this way of representing the boundary of
$\Delta$ is schematic, in the sense that the distance to the
boundary depends both on the choice of the origin and on direction.]
Because the entire bubble spacetime and its contents evolve from
data on CD, the entire bubble interior is completely determined [at
the classical level] by the properties of $\Delta$ and its contents.
The ``global point of view" criticised by advocates of
complementarity denies this and, in effect, attempts to describe
multiple copies of $\Delta$ simultaneously\footnote{Note however
that it has been suggested that the global and local points of view
might themselves be complementary: see \cite{kn:boucomp}.}.

The connection with the \emph{horizon} is as follows. The domain
$\Delta$ is not spherical in shape. The horizon of a given
observer must fit inside $\Delta$, with no protrusion beyond its
boundary [since such a protrusion would entail a violation of
complementarity if the horizon were to intersect horizons
contained in other copies of the fundamental domain]. This gives
an upper bound on the size of the horizon relative to $\Delta$.
However, it would be very unnatural if the horizon radius proved
to be \emph{much} smaller than the maximum. Ideally, we should be
able to \emph{prove} that $\Delta$ is \emph{just barely} able to
contain the horizon: in other words, we should find that $\Delta$
is positioned around the horizon in such a manner that the latter
appears as one of the largest possible balls in $\Delta$ which do
not protrude beyond the boundary. [In terms of the
compactification, this is formulated by saying that the ball
should be close to being as large as possible without
self-intersection.] That is, horizon complementarity should be on
or close to the brink of being violated. This is in agreement with
the principle put forward recently by Sekino and Susskind
\cite{kn:scrambler}, who argue that [black hole] complementarity
should ``just barely escape inconsistency": there should be no
``overkill". The situation is symbolized in Figure 1 by the
placing of the dot \emph{just outside} the cosmological horizon
[represented by the diagonal dashed line] of the observer at the
origin\footnote{We do not place the dot \emph{on} the horizon
because the the distance to the boundary of $\Delta$ is
direction-dependent. The situation shown represents a generic
direction.}.

Now that we have a general picture of the kind of overall geometry
for which we are aiming, we can try to be more specific. We begin
with the observation that if we take care to set up spatial boundary
conditions of physical fields in the fundamental domain $\Delta$, in
such a way that the description \emph{excludes all reference to
anything outside $\Delta$}, then there will be physical
consequences: we can expect the \emph{Casimir effect} to arise
\cite{kn:bytsenko}\cite{kn:noddy}. The Casimir effect arising when
negatively curved spaces are compactified in the manner proposed
here has been studied extensively: see for example
\cite{kn:helio}\cite{kn:mark}\cite{kn:neupane}\cite{kn:hoss}.
Typically, the Casimir effect has a local manifestation in terms of
an energy density, which definitely couples to gravity
\cite{kn:fulling}, and which \emph{violates the Null Energy
Condition} or NEC. This property of the Casimir energy plays a
crucial role here, since violations of the NEC often lead to
instabilities of various kinds, and the need to avoid these
instabilities yields useful constraints. The role of the Casimir
effect in shaping the overall structure of the bubble spacetime is
the theme of Section 3.

In order to be completely precise about this spacetime structure, we
need to know the value of the Casimir coupling to gravity, specified
here by a constant denoted by $\alpha$. In principle, this can be
worked out from the precise geometry of $\Delta$ and the detailed
structure of the matter fields present at the relevant time. In
practice this is an extremely difficult computation: even for
compact \emph{flat} spatial sections, which have a vastly simpler
geometry, the computations involved are very
intricate\footnote{There is an extensive literature on the use of
the Casimir effect in cosmological models with compact flat spatial
sections: see for example
\cite{kn:oddy}\cite{kn:coule}\cite{kn:szydgod}\cite{kn:levin}\cite{kn:sahara}\cite{kn:sahara1}\cite{kn:sahara2}.}.
We shall however argue that \emph{horizon complementarity may allow
us to compute the value of this parameter} in a very simple way, as
we shall shortly explain.

To proceed further, however, we need a more explicit description of
the relative positions of the various objects portrayed in Figure 1.
First, we need to specify the ``size" of $\Delta$, so that we can
control the horizontal position of the heavy dot. Second, we need to
know the [conformal] \emph{time} at which the bubble spacetime
emerges from the wall, relative to the [conformal] time
corresponding to the horizontal line ED at the top of the diagram.

Regarding the size of $\Delta$: we begin with the observation that
hyperbolic space has \emph{two} distinct length scales: one defined
by the curvature, and the other by the volume of the smallest
possible identical pieces into which it can be broken. The precise
way in which H$^3$ should be broken up into minimal-volume ``pieces"
was settled recently in a major work due to Gabai et al.
\cite{kn:gabai}, who proved that the well-known \emph{Weeks
manifold}, a certain compactified version of three-dimensional
hyperbolic space H$^3$, is the compactification with the smallest
possible volume. The maximum possible radius of a ball in the Weeks
manifold which does not self-intersect yields the second
distinguished length scale for H$^3$.

As was mentioned earlier, there are good physical reasons to suppose
that quantum-gravitational processes might favour spacetimes with
compact negatively curved spatial sections. This is in agreement
with the intuition \cite{kn:lindetypical} that it should be ``easier
to create" small universes than large ones. Following this to its
logical conclusion implies that, among the compact negatively curved
three-dimensional spaces, the ones with least volume should be
favoured. Furthermore, there is growing evidence \cite{kn:gabby} for
Thurston's long-standing conjecture that there is a precise
relationship between the combinatorial/topological \emph{complexity}
of a compact hyperbolic manifold and its volume, and it is
reasonable to argue
\cite{kn:twamley}\cite{kn:gibneg1}\cite{kn:gibneg2} that
quantum-gravitational effects favour low complexity. Thus we are led
to postulate that $\Delta$ is a fundamental domain for a Weeks
manifold.

The answer to our second question, regarding the time represented by
the surface CD, is less clear. However, there are two natural
``landmarks" in the evolution of this spacetime, and it seems
natural to assume that CD corresponds to one of them.

First, we shall see that there is a unique \emph{minimal} spacelike
hypersurface in the bubble spacetime; it owes its existence to the
Casimir effect. It seems natural to assume that this is the surface
along which a semi-classical description of the bubble interior
first becomes appropriate; one might regard this as another
application of the idea that ``small" spatial sections are favoured.
More physically, it has often been suggested
\cite{kn:york}\cite{kn:carlip} that the trace of the extrinsic
curvature of spatial sections is the true measure of the passage of
time in the earliest universe\footnote{The fundamental importance of
this quantity has also become apparent in recent deep analyses
\cite{kn:uggy} of the nature of spacetime singularities.}. If this
is true, then it is natural to suppose that the beginning of time
should correspond to the vanishing of this ``York time", and of
course this corresponds to the minimal surface here. Finally,
T-duality in string theory does suggest that the contracting part of
any ``bouncing" spacetime is probably a redundant description of
some other, expanding part of the spacetime, so that zero York time
is again favoured as the natural point for a semi-classical
description to be appropriate.

There is however an alternative to this proposal. We shall take
the matter content of the [earliest] bubble universe to consist of
both an inflaton field and the Casimir energy. Because of the
negative spatial curvature, the corresponding spacetimes contain
two spacelike hypersurfaces along which the energy densities
exactly cancel, so that the total energy of the universe vanishes
at those times. One can certainly argue \cite{kn:arrow} that a
semi-classical description should be established at such a time;
furthermore, cutting off the spacetime at the second of these two
surfaces has the advantage that the total energy of the bubble
universe will then \emph{never be negative}.

We shall consider both candidates for $\tau_{\m{e}}$, the conformal
time at which the semi-classical spacetime emerges from the bubble
wall: that is, we shall examine the consequences of identifying the
surface CD in Figure 1 with either the minimal surface or the
[second] zero-total-energy surface.

These assumptions regarding $\Delta$ and $\tau_{\m{e}}$ are
certainly the simplest one can make. The real test of their validity
is however the predictions they can generate, and this is the main
topic of the remainder of this work.

A direct application of horizon complementarity, as we have
interpreted it here, imposes a strong constraint on the shape of the
Penrose diagram for the ``Inflation plus Casimir" spacetime
discussed in section 3. This results in a \emph{lower} bound for the
Casimir parameter $\alpha$. This is discussed in Section 4.

The Casimir effect entailed by horizon complementarity violates the
Null Energy Condition [NEC]. In string theory, violations of the NEC
often lead to a particular kind of instability, discovered by
Seiberg and Witten \cite{kn:seiberg}. By requiring that the system
should not be unstable in this sense, and again using our
assumptions regarding $\tau_{\m{e}}$, we obtain an \emph{upper}
bound for $\alpha$. This computation is the subject of Section 5.

Thus we see that horizon complementarity leads to both an upper and
a lower bound on $\alpha$. In fact, the allowed range for $\alpha$
is extremely narrow, particularly in the case in which
$\tau_{\m{e}}$ = 0. This is very remarkable, because although both
constraints derive ultimately from horizon complementarity, one of
them depends on purely geometric properties of the Weeks manifold
and the other does not. Thus, in effect, we obtain a surprisingly
precise prediction for the value of $\alpha$, by means of two
applications of horizon complementarity. This also confirms that the
horizon just barely fits inside $\Delta$, in agreement with our
claim that the restriction to $\Delta$ accurately represents horizon
complementarity, and with the Sekino-Susskind ``no overkill"
principle \cite{kn:scrambler}. All this is discussed, with our
conclusions, in Section 6.

\addtocounter{section}{1}
\section* {\large{\textsf{3. The Role of the Casimir Effect}}}
In this section, we shall introduce a simple explicit spacetime
geometry, arising when the effects of Casimir energy are
superimposed on the geometry of an inflating Coleman-De Luccia
bubble universe. For the sake of clarity, let us recall that
geometry.

The [simply connected version of] global de Sitter spacetime with
[in the signature we use here] spacetime curvature 1/L$^2$ is
defined as the locus, in five-dimensional Minkowski spacetime
[signature ($-\;+\;+\;+\;+$)], defined by the equation
\begin{equation}\label{eq:A}
\m{-\; A^2\; + \;W^2 \;+ \;Z^2\; +\; Y^2\; + \;X^2\; =\; L^2}.
\end{equation}
This locus has topology $\bbr\times\,\m{S}^3$, and it can be
parametrized by \emph{global} conformal coordinates
($\eta,\,\chi,\,\theta,\,\phi)$ defined by
\begin{eqnarray} \label{eq:B}
\m{A} & = & \m{L\;cot(\eta)  }                     \nonumber \\
\m{W} & = & \m{L\;cosec(\eta)\;cos(\chi)}                     \nonumber \\
\m{Z} & = & \m{L\;cosec(\eta)\;sin(\chi)\;cos(\theta)}           \nonumber \\
\m{Y} & = & \m{L\;cosec(\eta)\;sin(\chi)\;sin(\theta)\;sin(\phi)}  \nonumber \\
\m{X} & = & \m{L\;cosec(\eta)\;sin(\chi)\;sin(\theta)\;cos(\phi)}.
\end{eqnarray}
Here $\chi,\,\theta,\,\phi$ are the usual coordinates on the
three-sphere, and $\eta$ is angular conformal time, which takes its
values in the interval ($0,\;\pi$). The metric of Global de Sitter
spacetime is then
\begin{equation}\label{eq:C}
\m{g(GdS)\; =\; L^2\,cosec^2(\eta)[ -\; d\eta^2 \; +\; d\chi^2 \;+\;
sin^2(\chi)\{d\theta^2  \;+\; sin^2(\theta)d\phi^2\}}].
\end{equation}
An obvious conformal transformation allows us to extend the range of
$\eta$, so that it takes all values in the closed interval
[$0,\;\pi$]. The Penrose diagram is clearly square [in the case of
\emph{simply connected} spatial sections], since $\chi$ also has
this range.

Now notice that the defining formula (\ref{eq:A}) is invariant under
an exchange, followed by a simultaneous complexification\footnote{It
is convenient to rotate A and W in opposite directions.}, of A and
W; so this transformation cannot change the local geometry.
Therefore, if we perform the exchange and complexify both $\eta$ [to
complexify A and W] and $\chi$ [so as then to avoid complexifying X,
Y, and Z], the resulting coordinates, defined by
\begin{eqnarray} \label{eq:BDS}
\m{A} & = & \m{-\,L\;cosech(\tau)\;cosh(\sigma)  }                     \nonumber \\
\m{W} & = & \m{L\;coth(\tau)}                     \nonumber \\
\m{Z} & = & \m{L\;cosech(\tau)\;sinh(\sigma)\;cos(\theta)}           \nonumber \\
\m{Y} & = & \m{L\;cosech(\tau)\;sinh(\sigma)\;sin(\theta)\;sin(\phi)}  \nonumber \\
\m{X} & = &
\m{L\;cosech(\tau)\;sinh(\sigma)\;sin(\theta)\;cos(\phi)},
\end{eqnarray}
are still coordinates on a spacetime locally identical to de Sitter
spacetime. However, complexification will change the nature of the
coordinates; the periodic coordinates are replaced by coordinates
taking values in an infinite range. Thus the new conformal time
coordinate $\tau$ ranges from zero to infinity, as does the
coordinate $\sigma$ which replaces $\chi$. The effect of this is
actually to \emph{restrict} the domain of these new coordinates:
they cannot cover the entire spacetime, because global de Sitter has
compact spatial sections. Comparing the expressions for A in
($\ref{eq:B}$) and ($\ref{eq:BDS}$), we see that $\eta\,> \, \pi/2$
on the domain of these coordinates, and then a comparison of the two
expressions for W shows that
\begin{equation}\label{eq:E}
\eta \;>\;{{\pi}\over{2}}\;+\;\chi.
\end{equation}
We see that the new coordinates actually parametrise only one-eighth
of the full Penrose diagram, the triangular top left-hand corner
extending upwards from the point $\chi$ = 0, $\eta$ = $\pi$/2.

This is, however, all we need to describe the interior of the bubble
portrayed in Figure 1, in the limit of an \emph{infinitely thin}
bubble wall [which corresponds to the null line in the conformal
diagram above which the ($\tau,\,\sigma$) coordinates are valid].
Thus it is reasonable to describe the region of de Sitter spacetime
covered by these coordinates as ``Bubble de Sitter". The metric in
these coordinates is
\begin{equation}\label{eq:H}
\m{g(BdS)\; =\; L^2\,cosech^2(\tau)[ -\; d\tau^2 \; +\; d\sigma^2
\;+\; sinh^2(\sigma)\{d\theta^2  \;+\; sin^2(\theta)d\phi^2\}}].
\end{equation}
We see at once that this piece of de Sitter spacetime is foliated by
spacelike hypersurfaces of constant \emph{negative} curvature. One
sees this also if one uses coordinates (t, r, $\theta$, $\phi$)
based on proper time: the same metric is now
\begin{eqnarray}\label{eq:G}
\m{g(BdS})\; =\;
\m{-\,dt}^2\;+\;\m{sinh^2\big(t/L\big)}\,\Big[\mathrm{dr^2\;
+\;\m{L^2}\, sinh^2(r/L)}\{\mathrm{d}\theta^2 \;+\;
\mathrm{sin}^2(\theta)\,\mathrm{d}\phi^2\}\Big];
\end{eqnarray}
here r = $\sigma$L. This replacement of the original spherical
spatial sections by hyperbolic sections is well-known, and its
importance has been particularly stressed by Freivogel et al.
\cite{kn:frei}, as a prediction of the Landscape. Less appreciated,
however, is that the complexification leading to this result
\emph{also renders conformal time infinite}. That is, $\tau$ runs
from 0 [corresponding to t = $\infty$] to $\infty$ [as t tends to
0].

We have arrived at this conclusion regarding the range of conformal
time inside the bubble because we have been studying the limiting
situation in which the bubble wall is infinitely thin, as explained
above. If we consider the more realistic situation portrayed in
Figure 1, with a thick bubble wall, then extremely large values of
$\tau$ are cut off along the surface CD, corresponding to one or the
other of the fixed ``landmarks" described in the preceding section.
Thus, the conformal diagram representing the semi-classical bubble
interior will be rectangular, with the conformal time of emergence
from the wall at the bottom. The corresponding value of $\tau$ will
not be \emph{infinitely} large. One should expect, however, that the
diagram will be ``very tall", in the sense that the range of $\tau$
will still be far larger than that of $\sigma$, if we restrict the
latter's domain by compactifying the spatial sections. That would be
unacceptable, since it is clear from Figure 1 that in fact the range
of $\sigma$ should [slightly] \emph{exceed} the range of conformal
time in the bubble.

On the other hand, we have also been assuming that the only energy
density that is relevant here is the one contributed by the inflaton
[in its potential-dominated state, so that it resembles a positive
vacuum energy]. In reality, other forms of energy might be present,
particularly in the region of the bubble wall ---$\,$ that is,
precisely in the region of large $\tau$. Can such additional matter
have the effect of ``shortening" the bubble's conformal diagram, as
we need?

To answer this, we need the beautiful theorem due to Gao and Wald
\cite{kn:gaowald}, which has a direct bearing on this question [see
also \cite{kn:gregnull}].

\bigskip
\noindent \textsf{THEOREM [Gao-Wald]: Let M be a spacetime
satisfying the Einstein equations and the following conditions:}

\noindent \textsf{[a] The Null Energy Condition [NEC] holds.}

\noindent \textsf{[b] M is globally hyperbolic and contains a
compact Cauchy surface.}

\noindent \textsf{[c] M is null geodesically complete and satisfies
the null generic condition.}

\noindent \textsf{Then there exist Cauchy surfaces S$_1$, S$_2$,
with S$_2$ $\subset$ I$^+$[S$_1$], such that, for any p $\in$
I$^+$[S$_2$], one has S$_1$ $\subset$ I$^-$[p].}

\bigskip

Here the null generic condition is the requirement that, along every
null geodesic, there should exist a point where the tangent vector
k$^{\m{a}}$ and the curvature R$_{\m{abcd}}$ satisfy
$\m{k_{[a}\,R_{b]cd[e}\,k_{f]}\,k^c\,k^d \neq 0}$, and I$^+$, I$^-$
denote respectively the chronological future [past] of an event or
set of events; see \cite{kn:waldbook}, Chapter 8. The Null Energy
Condition or NEC is the demand that the [full]
stress-energy-momentum tensor should satisfy
\begin{equation}\label{I}
\m{T_{ab}\,n^{a}\,n^{b}\;\geq\;0}
\end{equation}
at all points in spacetime and for all null vectors $\m{n^{a}}$.

In simple language, the Gao-Wald theorem means that a sufficiently
long-lived observer will, under the stated conditions, ultimately be
able to ``see" \emph{an entire spatial slice of the spacetime}. An
even simpler way of thinking about the theorem is as follows. Take
simply connected global de Sitter spacetime, and note that the
conclusion of this theorem is \emph{not} true of it. This is because
de Sitter spacetime is so ``special" that it does not actually
satisfy the null generic condition. In fact, because the conformal
diagram is square, global de Sitter spacetime \emph{just barely}
escapes having a ``fully visible" spatial section. Now suppose that
one introduces into this spacetime a small amount of homogeneously
distributed radiation. Then one can show explicitly
\cite{kn:tallandthin} that the effect is to cause the conformal
diagram to become ``\emph{taller}" [if we fix the width]. Of course,
this affects all parts of the diagram, so even if we decide to cut
off part of the spacetime at a finite time, the remaining part of
the diagram is also stretched vertically when radiation is
introduced. The Gao-Wald theorem states that this happens quite
generally: \emph{generic matter that satisfies the NEC makes the
conformal diagram of a spatially compact asymptotically de Sitter
spacetime grow taller.}

We see that the mere inclusion of ``other forms of energy" does not
help us; in fact, ``normal" matter just makes the situation worse.
The simplest way to circumvent the Gao-Wald theorem is to
\emph{violate the NEC}. We need to find a natural way of doing this
in the immediate vicinity of the bubble wall, without interfering
with the subsequent Inflation. We shall now argue that our proposal
---$\,$ to impose a formal compactification of the spatial sections
inside the bubble ---$\,$ automatically supplies the needed
mechanism.

Horizon complementarity requires us to give a complete description
of the physics inside the [inflationary] horizon, using \emph{only}
those fluctuations which are completely contained in a fundamental
domain $\Delta$, of the kind discussed in the preceding section.
Thus, complementarity requires us to impose boundary conditions of
precisely the kind which lead to the \emph{Casimir effect}
\cite{kn:bytsenko}\cite{kn:noddy}. This effect can be represented in
a formal way by including a negative energy component, which is
however non-negligible only when the universe is extremely small. As
is well known, the Casimir effect violates the NEC, and so it
provides us with precisely what we want: a natural, indeed
inevitable, way of evading the Gao-Wald theorem.

Let us set up a simple Friedmann model of this system; in doing so,
we are as usual ignoring the back-reaction of the inhomogeneities in
the Casimir energy distribution. [In the case of the Weeks manifold
with which we are concerned here, the relative inhomogeneities of
the Casimir energy tend in any case to be very mild; see
\cite{kn:helio}.] We shall work with the usual FRW spacetime
geometry, with scale factor a(t), where t is proper time, and with
negatively curved spatial sections on which we continue to use the
coordinates (r, $\theta$, $\phi$). [The fact that these sections
have been compactified is not apparent in the form of the local
metric; it is reflected only in the range of r, assuming that we
take this coordinate to be single-valued. See below.] Various kinds
of physical fields and compactification schemes contribute to the
Casimir energy in various ways and with different signs; we shall
assume that the total is \emph{negative} and depends on the inverse
fourth power of the scale factor
---$\;$ see for example \cite{kn:levin}. This is to be combined with
a positive vacuum energy density
$\rho_{\m{inflaton}}\,=\,+\,3/8\pi$L$^2$, representing the inflaton
in its potential-dominated state; here L is the typical inflationary
length scale [that is, the spacetime curvature at the end of
Inflation is 1/L$^2$], and we are using Planck units.

It will be convenient to represent the Casimir coupling to gravity
in the following way. Let us declare that the ratio of the magnitude
of the Casimir energy density to that of the inflaton is given by a
dimensionless positive constant $\gamma$ at the time when the scale
factor is equal to unity [assuming for the moment that this actually
occurs]. Thus, the Casimir energy density $\rho_{\m{casimir}}$ is
given by $-\,3\gamma/8\pi$L$^2$a$^4$. The Friedmann equation is
\begin{eqnarray}\label{eq:J}
\m{L^2\,\dot{a}^2\;=\;{{8\pi}\over{3}}\,L^2\,a^2\Big[\rho_{\m{inflaton}}\,+\,
\rho_{\m{casimir}}\Big]\;+\;1\;=\;{{8\pi}\over{3}}\,L^2\,a^2\Big[{{3}\over{8\pi
L^2}}\;-\;{{3\gamma}\over{8\pi L^2\,a^4}}\Big]\;+\;1.}
\end{eqnarray}

The solution for the ``Bubble de Sitter plus Casimir" metric [with
the constant of integration absorbed into the time coordinate] is
\begin{eqnarray}\label{eq:K}
\m{g(BdS\alpha}) = \m{-\;dt^2 + \Big[
\alpha^2+(1\,+\,2\alpha^2)\,sinh^2(t/L)\Big]\Big[dr^2 + \m{L^2}\,
sinh^2(r/L)}\{\mathrm{d}\theta^2 +
\mathrm{sin}^2(\theta)\,\mathrm{d}\phi^2\}\Big],
\end{eqnarray}
where
\begin{equation}\label{eq:L}
\alpha^2\;=\;\sqrt{{1\over 4}\;+\;\gamma}\;-\;{1\over 2}.
\end{equation}
Note that $\alpha$ is the smallest possible value of the scale
factor. This gives the geometric meaning of $\alpha$: if for example
$\Delta$ is a fundamental domain for the Weeks manifold, then [see
the next section] the minimal volume of a spatial section in this
spacetime is approximately 0.9427$\alpha^3$L$^3$, where L is the
inflationary length scale. We can now also clarify the physical
meaning of $\alpha$, as follows. Recall that $\gamma$ was defined as
a ``coupling" which measures the relative Casimir and inflaton
energy densities at the time when the scale factor equals unity.
However, a(t) = 1 has no particular significance in this problem
[and indeed it may not occur at any time], so the physical meaning
of $\gamma$ is obscure. We now see, however, that the maximal
magnitude of the Casimir energy density [which is clearly attained
at a unique time, t = 0] is
$|\rho^{\m{max}}_{\m{casimir}}|\,=\,3\gamma/8\pi$L$^2\alpha^4$, and
in fact one has
\begin{equation}\label{eq:LL}
\m{|\rho^{max}_{casimir}|\;=\;{3\over8\pi
L^2}\big(1\;+\;\alpha^{-\,2}\big)}.
\end{equation}
Thus $\alpha$ can be thought of as a number which parametrises the
maximum intensity of the Casimir effect in this problem. This
provides the physical meaning of this parameter. Henceforth we shall
refer to $\alpha$ as the ``Casimir coupling", since it is both
geometrically and physically more meaningful than $\gamma$. [In
fact, $\gamma\,=\,\alpha^2\,+\,\alpha^4$, so we can regard $\gamma$
as this function of $\alpha$.]

Note that the maximal value of the magnitude of the Casimir energy
density is, from (\ref{eq:LL}), always somewhat larger than the
inflaton energy density [provided that the hypersurface t = 0 is
indeed retained as part of the spacetime geometry ---$\,$ see
below]. Thus, the \emph{total} energy density is negative near to t
= 0. As the Casimir energy is diluted, the total energy density
passes through zero, when the scale factor is equal to
$\gamma^{1/4}$ [though the total \emph{pressure} is strictly
negative even at that time] and is thereafter positive. The formal
``total equation of state parameter" [the ratio of total pressure to
total energy density] is given by
\begin{equation}\label{eq:LLL}
\m{w_{tot}\;=\;{-\,(1\;+\;\gamma/3a^4)\over (1\;-\;\gamma/a^4)}},
\end{equation}
where a(t) is the scale factor. Notice that, immediately after the
time at which the total energy density vanishes, w takes on
arbitrarily large negative values. [It then rapidly approaches
$-\,1$.] This observation is of some interest in the light of the
recent work of Steinhardt and Wesley \cite{kn:steinhardt}, who argue
that Inflation can only arise in high-dimensional theories if w does
assume such values; that is, if the system is ``deep in the
NEC-violating regime". We see that Inflation \emph{automatically
satisfies this condition} if the Casimir effect is
present\footnote{Steinhardt and Wesley show that many other
conditions need to be satisfied; we are \emph{not} claiming that our
observation discussed here settles all of their concerns. [Nor does
the fact that they assume flat spatial sections mean that similar
results cannot be proved for bubble interiors; that remains to be
seen.]}.

It will be useful for us to write our metric in terms of
[dimensionless] conformal time, $\tau$, which is given by
\begin{equation}\label{eq:M}
\m{\tau\;=\;{1\over \alpha}\int_0^{t/L}{dt/L\over
\sqrt{1\;+\;[2\;+\;(1/\alpha^2)]sinh^2(t/L)}}\;=\;{- i \over
\alpha}\,F\Bigg({i t \over L}\;,\;\sqrt{2\;+\;(1/\alpha^2)}\Bigg)}.
\end{equation}
Here F($\phi \;,\;$ k) is the incomplete elliptic integral of the
first kind \cite{kn:abramo}, with Jacobi amplitude $\phi$ and
elliptic modulus k; in this case it has been evaluated along the
imaginary axis.

Inverting the elliptic integral we can express t in terms of the
amplitude:
\begin{equation}\label{eq:N}
\m{it/L\;=\;am\Big(i\alpha\tau \;,\;\sqrt{2\;+\;(1/\alpha^2)}\Big)}.
\end{equation}
Taking the sine of both sides we find that
\begin{equation}\label{eq:O}
\m{i\,sinh(t/L)\;=\;sn\Big(i\alpha\tau\;,\;\sqrt{2\;+\;(1/\alpha^2)}\Big)},
\end{equation}
where sn(u, k) is one of the classical Jacobi elliptic functions.
Using the formulae for complex arguments of elliptic functions given
on page 592 of \cite{kn:abramo}, one can express the right side as a
function of a real variable; substituting the result for sinh(t/L)
in equation (\ref{eq:K}) we obtain finally [with $\sigma$ = r/L]
\begin{eqnarray}\label{eq:P}
\m{g(BdS\alpha}) &=& \m{ L^2\,\Bigg[
\alpha^2+(1\,+\,2\alpha^2)\,{sn^2\Big(\alpha\tau\;,\;i\,\sqrt{1\;+\;(1/\alpha^2)}\Big)\over
cn^2\Big(\alpha\tau\;,\;i\,\sqrt{1\;+\;(1/\alpha^2)}\Big)}\Bigg]\;}
\nonumber \\
& & \times\;\m{\Big[ -\; d\tau^2 \; +\; d\sigma^2 \;+\;
sinh^2(\sigma)\{d\theta^2 \;+\; sin^2(\theta)d\phi^2\}\Big]. }
\end{eqnarray}
Here cn(u, k) is another of the Jacobi elliptic functions. This
metric is to be compared with the Bubble de Sitter metric given in
equation (\ref{eq:H}); the metric here is conformally the same as
that metric along future infinity; it is asymptotically de Sitter.
[In fact, the spacetime geometry here is significantly different
from that of Bubble de Sitter spacetime \emph{only} for a short time
after the emergence from the bubble wall, since the Casimir energy
is diluted away very rapidly with the expansion.]

The geometry here differs from that of Bubble de Sitter in an
important way, however: the extent of conformal time is \emph{not}
infinite. Its extent is instead given by setting the elliptic
function
$\m{cn\Big(\alpha\tau\;,\;i\,\sqrt{1\;+\;(1/\alpha^2)}\Big)}$ equal
to zero. The zeros of this function are given [\cite{kn:abramo},
page 590] by
\begin{equation}\label{eq:Q}
\m{cn(K(k), k)\;=\;0,}
\end{equation}
where K(k) is the \emph{complete elliptic function of the first
kind}. Thus $\tau$ has a formal range between
\begin{equation}\label{eq:R}
\m{\pm\,\tau_{\infty}(\alpha)\;=\;\pm\,{1\over
\alpha}\,K\Big(i\,\sqrt{1\;+\;(1/\alpha^2)}\Big).}
\end{equation}
Using the relevant formula from page 593 of \cite{kn:abramo} one can
express this in terms of real variables:
\begin{equation}\label{eq:S}
\m{\pm\,\tau_{\infty}(\alpha)\;=\;\pm {1\over
\sqrt{1\;+\;2\alpha^2}}\,K\,\Bigg(\sqrt{{1\;+\;\alpha^2\over
1\;+\;2\alpha^2}}\Bigg).}
\end{equation}
However, the negative value here corresponds to proper time t
tending to $-\,\infty$, which is not correct; it should instead be
replaced by the conformal time $\tau_{\m{e}}$ at which the bubble
universe \emph{emerges} from the bubble wall. Our hypothesis,
explained in the preceding section, is that $\tau_{\m{e}}$ either
vanishes or corresponds to the moment when the total energy of the
universe is zero. We shall return to that later; for the moment, let
us note that since the function K(k) diverges as k tends to unity
[it increases monotonically from a value of $\pi$/2 at k = 0, and in
particular is finite at $1/\sqrt{2}$ ---$\,$ see \cite{kn:abramo},
page 592], $\tau_{\infty}(\alpha)$ can be arbitrarily large if
$\alpha$ is very small, or arbitrarily small if $\alpha$ is large;
see Figure 2.
\begin{figure}[!h]
\centering
\includegraphics[width=0.7\textwidth]{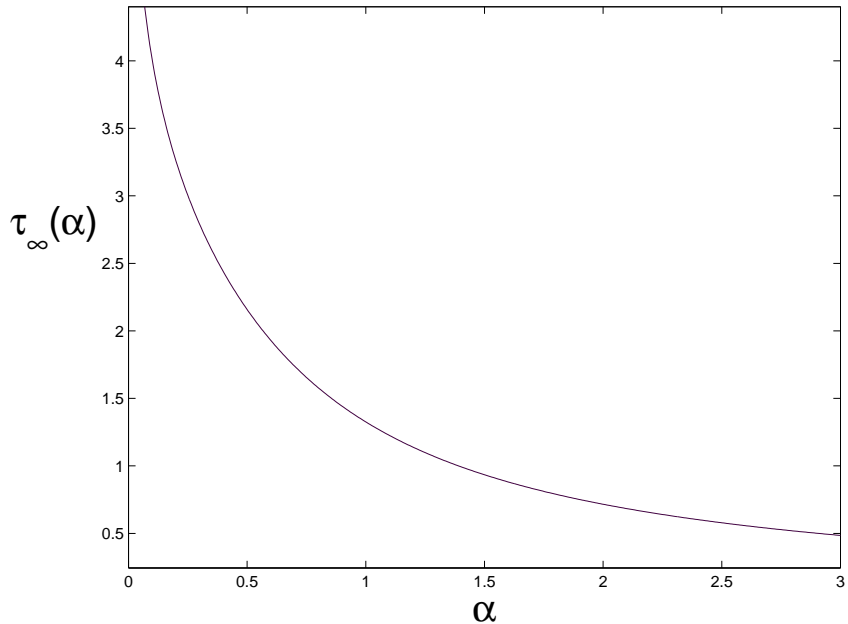}
\caption{Conformal Time to Infinity as a Function of $\alpha$.}
\end{figure}
Thus, we see that the shape of any conformal diagram for this
spacetime depends strongly on $\alpha$; and that, by adjusting
$\alpha$, we can avoid having a conformal diagram which is too tall.

This is a very satisfactory conclusion. We saw that the original
bubble de Sitter spacetime has a conformal diagram which is
extremely ``tall", and that the presence of matter satisfying the
NEC only made the situation worse. However, our proposed
interpretation of horizon complementarity automatically leads us to
include, in addition to the inflaton, another form of energy ---$\,$
Casimir energy ---$\,$ which allows us to evade the conclusions of
the Gao-Wald theorem and to restrict the range of conformal time in
just the way demanded by Figure 1. In the next section we will show
how to be more precise about this.

We see, then, that horizon complementarity leads to a rather
specific description of the spacetime geometry of the earliest part
of the history of a bubble universe. All that remains is to
determine the constant $\alpha$. We shall now argue that horizon
complementarity also constrains this parameter very strongly.

\addtocounter{section}{1}
\section* {\large{\textsf{4. The Geometry of the Weeks Manifold: First Constraint on $\alpha$}}}
It was shown long ago by Thurston \cite{kn:thursty} that there
exists a decomposition of hyperbolic three-space H$^3$ into
identical pieces of minimal volume ---$\;$ that is, that there is a
minimal-volume compactification. It was long conjectured, and
finally proved by Gabai et al. \cite{kn:gabai}, that this
distinguished decomposition corresponds to the \emph{Weeks
manifold}\footnote{A good description of the Weeks manifold, with
illustrations, may be found in \cite{kn:inoue}; see also
\cite{kn:gomero}.}, \textbf{W}. For this manifold, the fundamental
domain can be represented as a certain hyperbolic polyhedron with 18
faces. The volume is $\approx$ 0.9427$\times$$\lambda$$^3$, where
$\lambda$ is the curvature radius. Our hypothesis is that this is
the hyperbolic compactification which defines the domain $\Delta$.

Recall that we propose to relate the fundamental domain to the
cosmic horizon [at the time of emergence from the bubble wall] by
identifying the horizon as a sphere which is completely enclosed by
the domain. The systematic study of such spheres can be briefly
explained as follows.

In any compact manifold M, the \emph{injectivity radius} I(M, p) at
a point p in M is defined as the maximal radius of a sphere centred
at p which does not self-intersect. [That is, the maximal radius
such that the exponential map is injective.] For a sphere or a
torus, this quantity is actually independent of the point p, but
\emph{this is not so} for compact hyperbolic manifolds, for which
the boundary of a fundamental domain is much more irregular. That
is, the injectivity radius is a function of position on a compact
hyperbolic manifold.

The range of sizes of spheres which can be contained in a compact
hyperbolic space can be surveyed as follows. For each such space one
can define an \emph{injectivity distribution}, a function introduced
by Weeks \cite{kn:weeks} and defined as follows. Let dV/V be the
fraction of the volume of M containing points p with I(M, p)
[measured in units of $\lambda$] lying between the values x and x +
dx. Then the injectivity distribution is the function on the real
line defined by
\begin{equation}\label{eq:T}
\m{ID(M; x)\;=\;(dV/V)/dx.}
\end{equation}
That is, ID(M; x) measures the rate at which the fractional volume
containing points with a given injectivity radius changes with
increasing injectivity radius; integrating it between selected
values of x gives the fraction of the volume of M containing points
with injectivity radii between those values. The curve representing
ID(M; x) [which need not be a continuous function] intersects the x
axis at two points. The smaller of these two values signals the
radius at which it becomes possible for a sphere to self-intersect;
the larger signals the radius beyond which this must happen. This
latter quantity is sometimes called the \emph{inradius} of M, which
we denote by $\sigma$(M). [Here $\sigma$ refers to the dimensionless
radial coordinate used in equation (\ref{eq:P}).]

The functions ID(M; x) are given in approximate form for ten
low-volume hyperbolic spaces in \cite{kn:weeks}. In particular, for
the Weeks manifold \textbf{W} [``Manifold 1" in \cite{kn:weeks}],
ID(\textbf{W}; x) is a function which has support on an interval
extending roughly from 0.292 to 0.519 $\approx$
$\sigma$(\textbf{W}). Thus, a horizon of conformal radius less than
0.292 can be located anywhere in H$^3/\Gamma$ without danger of
intersecting itself; but a horizon of conformal radius larger than
0.519 would have to do so, no matter where it might be located.
[Note that this last quantity, the inradius, varies between roughly
0.5 and 0.6 for the ten low-volume manifolds examined in
\cite{kn:weeks}; however it can be substantially larger than this
for other well-known compact hyperbolic manifolds; it is
approximately 0.996 for the Seifert-Weber space, the most easily
visualised compact hyperbolic manifold \cite{kn:thurston}.]

From the above discussion of the injectivity distribution, one sees
that the spacetime with local metric given in (\ref{eq:P}) and
spatial sections of the form H$^3/\Gamma$ does not have a Penrose
diagram in the conventional sense, because the fundamental domain
$\Delta$ is neither [globally] rotationally symmetric nor [globally]
homogeneous. That is, if we take the metric in (\ref{eq:P}) to be
defined on H$^3/\Gamma$, then the range of the conformal coordinate
$\sigma$, while always finite if we demand that it be single-valued,
depends on position and orientation. Therefore, one has a collection
of diagrams, one for location in $\Delta$ and one for each
direction; each is rectangular, with the same height [for given
assumptions regarding the range of conformal time] but with varying
widths. This situation is represented by the ``Penrose diagram" in
Figure 3, where the dotted rectangles represent possible diagrams
for various directions and positions. The axes here correspond to
the conformal coordinates $\tau$ and $\sigma$ in (\ref{eq:P}); the
diagram represents the spacetime geometry when the overall conformal
factor has been removed.
\begin{figure}[!h]
\centering
\includegraphics[width=0.8\textwidth]{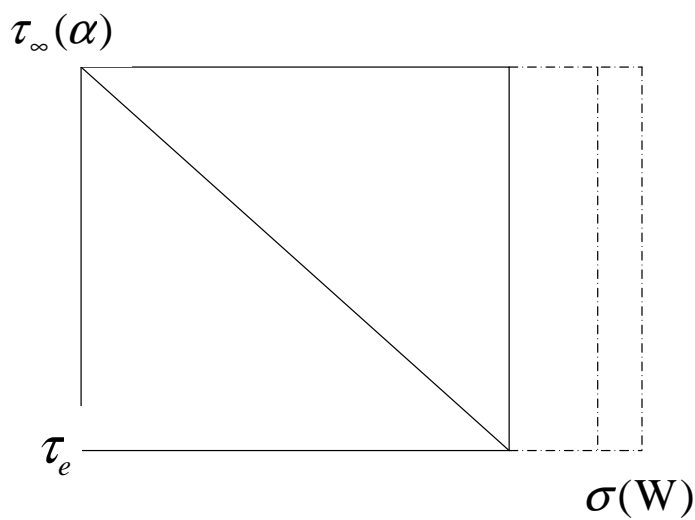}
\caption{``Penrose Diagram" for a Spatially Compactified Bubble.}
\end{figure}

The coordinate $\tau$ ranges from its value $\tau_{\m{e}}$ at the
time of emergence from the bubble wall up to
$\tau_{\infty}(\alpha)$. The coordinate $\sigma$ ranges from zero up
to a certain maximum; the extreme case of interest here corresponds
to a sphere located at a point where the injectivity radius is as
large as possible, taking the value $\sigma$(\textbf{W}). The
horizon is, as usual, the diagonal line in the figure. Since we
insist that the horizon must never, at any location in $\Delta$,
extend beyond the latter's boundary, we have a simple inequality:
\begin{equation}\label{eq:U}
\m{\tau_{\infty}(\alpha)\;-\;\tau_e\;<\;\sigma(\textbf{W}),}
\end{equation}
though, as we have explained, we would certainly prefer to replace
the inequality by an [approximate] equality.

We have two candidate assumptions for $\tau_{\m{e}}$. The first is
that it should be taken to be zero: we accept exactly the expanding
half of the original spacetime. In that case, we have
\begin{equation}\label{eq:UU}
\m{\tau_{\infty}(\alpha)\;<\;\sigma(\textbf{W})\;\approx\;0.519.\phantom{aaaaaaaaa}[\tau_e
= 0]}
\end{equation}
A numerical investigation reveals that equation (\ref{eq:S}) now
implies
\begin{equation}\label{eq:UUU}
\m{\alpha\;>\;2.80,\phantom{aaaaaaaaa}[\tau_e = 0]}
\end{equation}
approximately. This can be checked by inspecting the graph in Figure
2. Note that because the graph is quite flat for values of $\alpha$
beyond about 2, the replacement of the Weeks manifold by a manifold
[if one existed] with even a slightly smaller inradius would
strengthen this constraint significantly; on the other hand, its
replacement by [say] the Seifert-Weber space [inradius approximately
0.996] would weaken the constraint on $\alpha$ somewhat less
dramatically.

The alternative hypothesis is that $\tau_{\m{e}}$ corresponds to the
conformal time $\tau_{\rho \,=\, 0}$ when the total energy density
of the universe is zero. This occurs when the scale factor increases
from its minimal value [given by a(0) = $\alpha$] to
\begin{equation}\label{eq:UUUU}
\m{a(t_{\rho \,=\, 0})\;=\;[\alpha^2\;+\;\alpha^4]^{1/4},}
\end{equation}
where t$_{\rho \,=\, 0}$ is the proper time corresponding to
$\tau_{\rho \,=\, 0}$. Using the scale factor as the variable, we
can show after a short calculation that (\ref{eq:UU}) must be
replaced by
\begin{equation}\label{eq:UUUUU}
\m{{1\over
\sqrt{1\;+\;2\alpha^2}}\,K\,\Bigg(\sqrt{{1\;+\;\alpha^2\over
1\;+\;2\alpha^2}}\Bigg)\;-\;\int_{\alpha}^{[\alpha^2\;+\;\alpha^4]^{1/4}}{da\over
\sqrt{a^4\,+\,a^2\,-\,\alpha^2\,-\,\alpha^4}}\;<\;\sigma(\textbf{W}),\phantom{a}[\tau_e
= \tau_{\rho \,=\, 0}]}
\end{equation}
where we remind the reader that K(k) is a certain elliptic function.

Although this is far from obvious, a rather more involved numerical
investigation shows that the complicated function of $\alpha$ on the
left side of this inequality is still a decreasing function; hence
(\ref{eq:UUUUU}) still yields a lower bound on $\alpha$. We find
that the constraint on $\alpha$ is now somewhat weaker:
\begin{equation}\label{eq:UUUUUU}
\m{\alpha\;>\;2.42.\phantom{aaaaaaaaa}[\tau_e = \tau_{\rho \,=\,
0}]}
\end{equation}

Of course, both (\ref{eq:UUU}) and (\ref{eq:UUUUUU}) still allow for
very large values of $\alpha$. That would correspond to horizon
radii which are very much smaller than they need to be in order to
fit inside the domain $\Delta$. Since the decomposition of the
spatial sections into copies of $\Delta$ is supposed to be a formal
description of horizon complementarity, that would be a
disappointing conclusion, and furthermore it would conflict with
Sekino and Susskind's ``no overkill" principle \cite{kn:scrambler};
it would be far preferable to find that the horizon just barely fits
inside $\Delta$. We shall now argue that this is, in fact, exactly
what happens.

\addtocounter{section}{1}
\section* {\large{\textsf{5. Non-Perturbative Stability: Second Constraint on $\alpha$}}}
As we have seen, it is essential for our purposes that the Casimir
effect violates the Null Energy Condition. It is well known,
however, that such violations often lead to serious instabilities.
We now consider this.

The status of the NEC has been much debated of late, from various
points of view [see for example
\cite{kn:od}\cite{kn:brandy}\cite{kn:crem} and references therein].
NEC violation may or may not be acceptable in cosmology
---$\,$ the recent work of Steinhardt and Wesley
\cite{kn:steinhardt} suggests that it may simply be
\emph{unavoidable} if Inflation is embedded in a higher-dimensional
theory ---$\,$ but it is certainly the case that there are many
circumstances in which it leads to major problems [due to ghosts and
gradient energies of the wrong sign
\cite{kn:dubovsky}\cite{kn:kal}]. Exceptions do arise, however, in
strictly quantum, non-local systems \cite{kn:nimah}.

One such exception occurs around a black hole which is not in
equilibrium with its own Hawking radiation \cite{kn:strominger}; in
that case, the NEC violation is associated with the ``quantum
defocussing" which allows the event horizon to contract. The second
main example of physically acceptable NEC violation is provided
precisely by the Casimir effect we have been discussing.
Arkani-Hamed et al. argue \cite{kn:nimah} that this intrinsically
quantum effect is acceptable because one cannot use it to construct
``non-gravitating clocks and rods."

However, while Casimir energy may be acceptable in most cases, this
is not to say that it is always completely innocuous. In particular,
since the work of Seiberg and Witten \cite{kn:seiberg}, it has been
clear that branes and other extended objects can lead to forms of
instability which are completely obscure in any perturbative
approach. [More complex kinds of non-perturbative instability are
also known \cite{kn:horpolch}.] It is not obvious that Casimir
energy is always harmless in the non-perturbative context.

\emph{Seiberg-Witten instability} refers to the uncontrolled
nucleation of branes in certain modified versions of anti-de Sitter
spacetime; it depends on a delicate interplay between the growth of
volumes and surface areas in asymptotically hyperbolic Euclidean
geometries\footnote{A Riemannian manifold is said to be
\emph{asymptotically hyperbolic} if it has a well-defined conformal
boundary. The methods introduced by Seiberg and Witten apply to any
spacetime with an asymptotically hyperbolic Euclidean version.}.
Unfortunately, there is no known general criterion which might allow
us to decide, without a detailed calculation, whether a given system
is free of Seiberg-Witten instabilities; normally one has to
evaluate the relevant action explicitly, and show for example that
it grows monotonically away from a given non-negative value.
Explicit examples of systems which are unstable in this sense were
discussed by Maldacena and Maoz \cite{kn:maoz}; see also
\cite{kn:porrati}\cite{kn:tallandthin}. The method of direct
examination of the action was also used recently to rule out AdS
black holes with certain exotic event horizon topologies
\cite{kn:conspiracy} and to verify the non-perturbative stability of
certain solutions of M-theory \cite{kn:marealle}.

While no completely general criterion for the occurrence of
Seiberg-Witten instability is known, many concrete examples show
that NEC violation does have a \emph{tendency} to induce it
\cite{kn:unstable}. Thus, we must take care to ensure that this does
not occur in our case. To put it another way: by \emph{requiring}
that it should not occur, we may obtain a useful constraint.

In four spacetime dimensions the brane action has, for BPS branes,
the general form
\begin{equation}\label{eq:V}
\mathrm{S} \;=\;
\Theta\Big(\mathrm{A}\;-\;{{\mathrm{3}}\over{\mathrm{L}}}\,\mathrm{V}\Big),
\end{equation}
where $\Theta$ is the tension, A is the brane area, V the volume
enclosed, and L is the background asymptotic curvature radius. It is
clear that this object can be \emph{everywhere} non-negative, as it
must be if Seiberg-Witten instability is to be avoided, only in a
very particular kind of geometry: one where the area grows extremely
rapidly with radius.

Let us see how this works in the simplest possible asymptotically
hyperbolic geometry, namely four-dimensional hyperbolic space H$^4$
itself. Here it is easy to show that the action is positive at
\emph{large} distances, but the situation is less clear at small
distances from the origin\footnote{Seiberg and Witten gave a simple
criterion [positivity of the scalar curvature of the boundary] for
the brane action to be positive at \emph{large} distances.}. The
metric here, using the global foliation of H$^4$ by three-spheres,
is
\begin{equation}\label{eq:W}
\m{g(H^4)\; =\; dt^2\;+\;L^2\,sinh^2\big(t/L\big)\,\Big[d\chi^2\;
+\;sin^2(\chi)\{d\theta^2 \;+\; sin^2(\theta)\,d\phi^2}\}\Big],
\end{equation}
and the brane action is
\begin{equation}\label{eq:X}
\m{S[H^4](\Theta,
\,L\,;\;t)\;=\;2\pi^2\,\Theta\,L^3\,\Big[sinh^3(t/L)\;-\;{{1}\over{4}}\,cosh(3t/L)\;+\;{{9}\over{4}}\,cosh(t/L)\;-\;2\,
\Big].}
\end{equation}
Note that the negative second term here is actually \emph{larger} in
magnitude than the first term for small values of t, underlining the
fact that the positivity of the action is somewhat precarious even
here, in the case of undisturbed ``pure" hyperbolic space.
Nevertheless, one can verify that [because of the presence of the
third term, which is negligible at large distances] this function,
which obviously vanishes at t = 0, is monotonically increasing, and
hence is everywhere non-negative; see Figure 4. [In this and in the
subsequent diagrams, the t-axis has units given by L; the units on
the vertical axis are given by $2\pi^2\,\Theta\,$L$^3$.]

\begin{figure}[!h]
\centering
\includegraphics[width=0.6\textwidth]{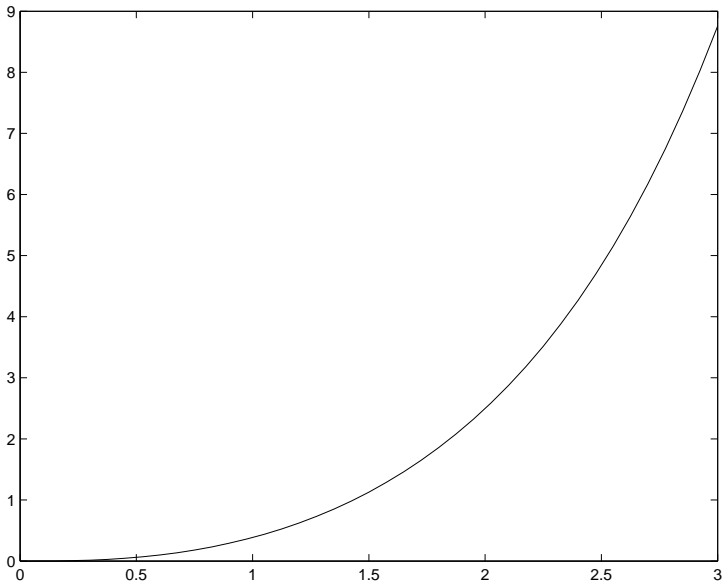}
\caption{Brane Action for Pure H$^4$.}
\end{figure}

Clearly, however, one would not need to modify the geometry of
hyperbolic space very drastically in order to cause the brane action
to become negative near to the origin. The Seiberg-Witten brane
action is in fact a much more subtle object than its simple
appearance might suggest.

One modification which certainly has a clear potential to lead to
trouble here is to introduce a ``wormhole-like" structure into
H$^4$, that is, some kind of structure such that a large volume is
contained in a small area; for this will allow the volume term in
(\ref{eq:V}) to dominate the area term in the vicinity of the
``wormhole-like" object. In general terms, there are strong
suggestions from several directions
\cite{kn:yau}\cite{kn:malda}\cite{kn:arkhole}\cite{kn:hsu2} that
such objects are in fact unacceptable in quantum gravity. The
requirement of Seiberg-Witten stability allows us to be more precise
in the case of interest to us.

The point is that the Euclidean geometry corresponding to our metric
(\ref{eq:K}) is indeed ``wormhole-like": to be precise, it
corresponds to \emph{part of} a Euclidean wormhole. [This is of
course not surprising, in view of the familiar association of
wormholes with NEC violation.] To see this, note that the metric
given in (\ref{eq:W}) is precisely the asymptotically hyperbolic
Euclidean version of the Bubble de Sitter metric (\ref{eq:G}): one
obtains (\ref{eq:W}) from (\ref{eq:G}) simply by complexifying t and
L and re-labelling r. Performing this same complexification on
(\ref{eq:K}), we obtain the asymptotically hyperbolic Euclidean
space with metric
\begin{eqnarray}\label{eq:Y}
\m{g(AHBdS\alpha)} &=& \m{dt^2} \;+\;  \m{L^2\,\Big[
\alpha^2\;+\;(1\;+\;2\alpha^2)\,sinh^2(t/L)\Big]}
\nonumber \\
& & \phantom{aaaaaaaaaa} \times\;\m{\Big[d\chi^2 \;+\;
sin^2(\chi)}\{\mathrm{d}\theta^2 \;+\;
\mathrm{sin}^2(\theta)\,\mathrm{d}\phi^2\}\Big].
\end{eqnarray}
If we allow t to run from $-\,\infty$ to $+\,\infty$, then this is
like two copies of H$^4$ joined smoothly near the origin by a
wormhole. But clearly the part of the space corresponding to
\emph{negative} values of t will lead to problems here: if that part
of the wormhole is retained, then we run the risk that some branes
at positive t will have small areas and large volumes, causing the
brane action to become negative. We see that non-perturbative
stability actually \emph{forces} us to truncate the manifold at some
value t = t$_{\m{e}}$ of the proper time, corresponding to the
emergence of a semi-classical bubble spacetime from the bubble wall.
In view of our discussion in the preceding section, this is very
satisfactory; we now have a method of assessing the physical
consequences of truncating the spacetime in either of the two ways
discussed in the previous section.

For the metric given in equation (\ref{eq:Y}), the brane action for
t $\geq$ t$_{\m{e}}$ is
\begin{eqnarray}\label{eq:Z}
\m{S[AHBdS\alpha](t_e,\,\Theta, \,L\,;\;t)} &=&
\m{2\pi^2\,\Theta\,L^3\,\Bigg[\,\Big(
\,\alpha^2\;+\;(1\;+\;2\alpha^2)\,sinh^2(t/L)\Big)^{3/2}\;} \nonumber \\
                        & & \phantom{aaa} - \;\m{{{3}\over{L}}\,\int_{t_e}^t\,\Big(
\,\alpha^2\;+\;(1\;+\;2\alpha^2)\,sinh^2(u/L)\Big)^{3/2}du\Bigg]},
\end{eqnarray}
where $\Theta$ is the tension, as in equation (\ref{eq:V}). As the
metric here is asymptotically indistinguishable from that of the
pure hyperbolic space discussed above, this function is certainly
positive at large t. The problem is to understand what happens at
small values of t, where the NEC violation is most intense.

The derivative of the action with respect to t can be expressed,
after a straightforward calculation, as
\begin{eqnarray}\label{eq:AA}
\m{{dS[AHBdS\alpha](t_e,\,\Theta, \,L\,;\;t)\over dt}} &=&
\m{3\pi^2\,\Theta\,L^2\,\Big[
\,1\;-\;(1\;+\;2\alpha^2)\,e^{-\,2t/L}\Big]} \nonumber \\
                        & & \phantom{aaaaaaaa} \times\;\m{\Big[
\,\alpha^2\;+\;(1\;+\;2\alpha^2)\,sinh^2(t/L)\Big]^{1/2}}.
\end{eqnarray}
Let us consider our first proposal for the truncation, along the
hypersurface of zero extrinsic curvature t$_{\m{e}}$ =
$\tau_{\m{e}}$ = 0; we return to the alternative, a truncation along
a surface of zero total energy density, below. We begin by stressing
that the initial value of the action is not zero for $\alpha > 0\,$:
it is equal to the positive value
$\m{2\pi^2\,\Theta\,L^3}\,\alpha^3$. Furthermore, one can show that
the second derivative is positive everywhere. On the other hand, we
see at once that the slope of the graph of the action is
\emph{negative} at t = 0; it is equal to $\m{-\,6
\pi^2\,\Theta\,L^2}\,\alpha^3$. The action function is \emph{not}
monotonically increasing, as it is in the case of pure hyperbolic
space; as we feared, there is a real possibility that the action
could become negative. It can be shown that this initial decrease of
the brane action is due to the fact that \emph{the Casimir effect
violates the NEC.} We see that NEC violation tends to induce
Seiberg-Witten instability, as claimed.

The graph of the action reaches a unique minimum [as can be seen
from (\ref{eq:AA})] at a positive value of t, namely t =
(ln($\sqrt{1 + 2\alpha^2}$))L. The system will be stable in the
Seiberg-Witten sense provided that the action is non-negative at
this point. That is, if we define a number $\Xi_{\alpha}$, depending
only on $\alpha$, by
\begin{eqnarray}\label{eq:BB}
\m{\Xi_{\alpha}\;=\;{S[AHBdS\alpha](t_e = 0,\,\Theta,
\,L\,;\;(ln(\sqrt{1 + 2\alpha^2}))L)\over 2\pi^2\Theta L^3}},
\end{eqnarray}
then the system is non-perturbatively stable if and only if
$\Xi_{\alpha}\,\geq\,0$.

In fact, a numerical investigation shows that the action \emph{does}
become negative if $\alpha$ is sufficiently large, showing that
Seiberg-Witten instability is a possibility here even in the absence
of half of the wormhole. For example, if we take $\alpha$ = 5, then
it is clear from Figure 5 that the system will be unstable in the
Seiberg-Witten sense.
\begin{figure}[!h]
\centering
\includegraphics[width=0.7\textwidth]{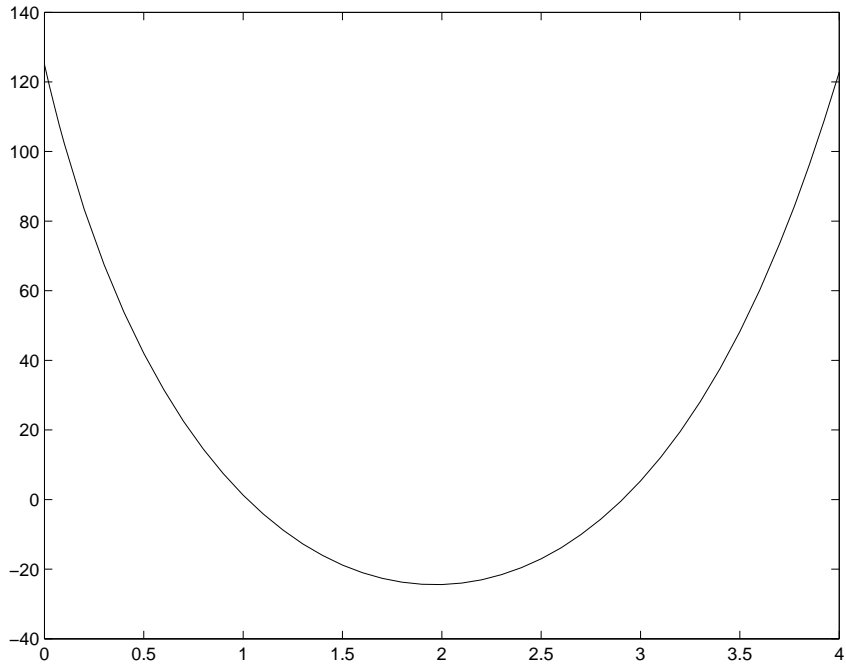}
\caption{Brane Action, $\alpha$ = 5, $\tau_{\m{e}}$ = 0.}
\end{figure}
If we think of $\Xi_{\alpha}$ as a function of $\alpha$, then we see
that this function is already negative at $\alpha$ = 5, and in fact
it becomes steadily more negative as $\alpha$ increases beyond 5. We
know that $\Xi_{\alpha}$ is zero at $\alpha$ = 0 [Figure 4], and it
would be perfectly reasonable to expect that it is negative for
\emph{all} positive $\alpha$; this would mean that NEC violation
leads to instability in all cases, which would not be entirely
unexpected.
\begin{figure}[!h]
\centering
\includegraphics[width=0.7\textwidth]{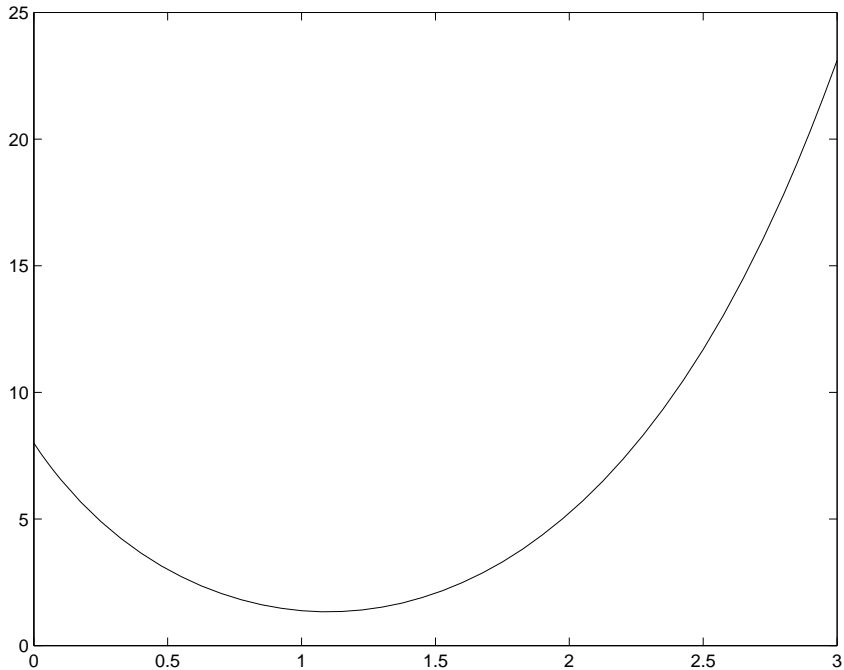}
\caption{Brane Action, $\alpha$ = 2, $\tau_{\m{e}}$ = 0.}
\end{figure}
Remarkably, however, numerical experiments show that this is
\emph{not} the case: as a function of $\alpha$, $\Xi_{\alpha}$ is
actually positive for a brief interval near to $\alpha$ = 0. For
example, it is positive at $\alpha$ = 2: see Figure 6. Further
experimentation shows that $\Xi_{\alpha}$ is approximately zero at
around $\alpha$ = 2.88, as shown in Figure 7.
\begin{figure}[!h]
\centering
\includegraphics[width=0.7\textwidth]{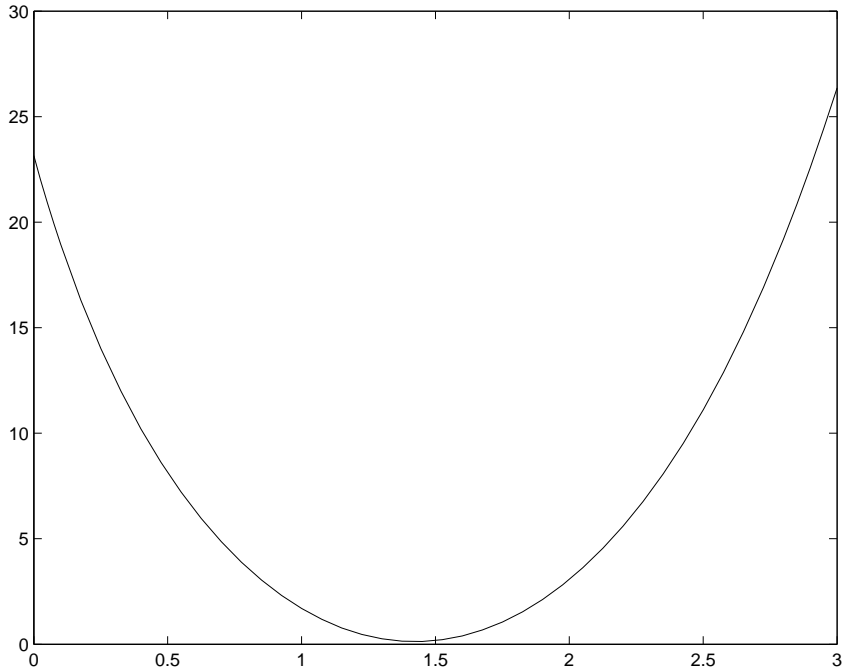}
\caption{Brane Action, $\alpha$ = 2.88, $\tau_{\m{e}}$ = 0.}
\end{figure}
Thus, despite the NEC violation associated with the Casimir effect,
the system is actually non-perturbatively \emph{stable} for values
of $\alpha$ between zero and about 2.88, but not beyond that value;
so we have
\begin{equation}\label{eq:CC}
\m{\alpha\;<\;2.88.\phantom{aaaaaaaaa}[\tau_e = 0]}
\end{equation}

Let us consider our alternative proposal for $\tau_{\m{e}}$, defined
now as the conformal time at which the total energy density
vanishes. This is a more complex situation computationally, because
now $\tau_{\m{e}}$ and t$_{\m{e}}$ are themselves functions of
$\alpha$, as one can see from equation (\ref{eq:UUUU}). Thus the
action given in equation (\ref{eq:Z}) acquires a still more
complicated dependence on $\alpha$. We find numerically [again, the
use of the scale factor as the variable is helpful] that requiring
the action to be non-negative in this case still imposes an upper
bound on $\alpha$, albeit a slightly weaker one: (\ref{eq:CC}) is
replaced by
\begin{equation}\label{eq:CCC}
\m{\alpha\;<\;2.91.\phantom{aaaaaaaaa}[\tau_e = \tau_{\rho \,=\,
0}]}
\end{equation}

We see that the zero-density definition of $\tau_{\m{e}}$ imposes
weaker conditions on $\alpha$, both from above and from below, than
the zero-extrinsic-curvature definition; though the effect on the
upper bound is much smaller than that on the lower bound.

\addtocounter{section}{1}
\section* {\large{\textsf{6. Conclusion: Using Horizon Complementarity to Predict $\alpha$}}}

We can summarize as follows. We have argued that horizon
complementarity should be implemented in the case of bubble
universes by means of the scheme represented by Figure 1. It turns
out that this requires the presence of some other form of energy,
apart from that of the inflaton, in the earliest history of the
bubble world; but fortunately the compactification itself supplies a
natural way of ensuring this, through the Casimir effect. Requiring
the conformal diagram to take a shape consistent with Figure 1
imposes a \emph{lower} bound on the gravitational Casimir coupling.
But violating the NEC forces us to impose an \emph{upper} bound on
the coupling if non-perturbative instability is to be avoided. With
our two natural candidates for the time of emergence of the bubble
universe from the bubble wall, we obtain combined constraints of the
form [see the inequalities
(\ref{eq:UUU})(\ref{eq:CC})(\ref{eq:UUUUUU})(\ref{eq:CCC})]
\begin{equation}\label{eq:ALPHA}
\m{2.80\;<\;\alpha\;<\;2.88,\phantom{aaaaaaaaa}[\tau_e = 0]}
\end{equation}
\begin{equation}\label{eq:BETA}
\m{2.42\;<\;\alpha\;<\;2.91.\phantom{aaaaaaaaa}[\tau_e = \tau_{\rho
\, = \,0}]}
\end{equation}
Inserting a value of $\alpha$ in one of these ranges into equation
(\ref{eq:K}), we have a completely definite form for the metric of
the bubble spacetime during its earliest history. We see that while
the compactification may only be enforced beyond the horizon, it
does strongly constrain the local metric: in fact, the metric is
nearly fixed.

The constraints we have obtained on $\alpha$ are remarkably tight,
astonishingly so in the $\tau_{\m{e}}$ = 0 case. However, the reader
is entitled to object that the precision of the resulting prediction
for $\alpha$, while very striking, is an illusion: clearly we have
made approximations in order to arrive at these results, and it is
not obvious how these can be quantified at this level of precision.

While there is justice in this observation, we wish to draw the
reader's attention to the following. Both the upper and the lower
bounds on $\alpha$ were in each case obtained from horizon
complementarity; but one has no other reason to expect them to
agree, even roughly. Recall that (\ref{eq:UUU}) and
(\ref{eq:UUUUUU}) are imposed directly by the geometry of the Weeks
manifold [in particular, by the value taken by its inradius]. Until
the recent work of Gabai et al. \cite{kn:gabai}, the assertion that
the Weeks manifold [volume approximately 0.9427 in curvature units]
is the ``smallest" compact hyperbolic space was little more than a
guess: the most recent actually established previous lower bound for
the volume of such a space \cite{kn:agol} was considerably lower, at
around 0.67. If a compact hyperbolic manifold with such a volume had
existed, its inradius would presumably have been somewhat smaller
than the inradius of the Weeks manifold [$\approx$ 0.519]. If the
inradius had been [say] 0.45, then for example (\ref{eq:UUU}) would
have been replaced by a requirement that $\alpha$ should be at least
3.24, which would not be easy to reconcile with (\ref{eq:CC}). In
short, even getting the ranges permitted by the two constraints to
overlap is not trivial. Indeed, if the precision of the agreement of
(\ref{eq:UUU}) and (\ref{eq:CC}) is mere coincidence, it remains
remarkable that the [pure] numbers involved are even \emph{similar
in order of magnitude.} This encourages the hope that a more
sophisticated treatment might lead to a reasonable estimate for
$\alpha$.

Ultimately it should be possible to check this claim directly. With
a more complete understanding, perhaps derived from string
cosmology, of the precise nature of the inflaton [and of whatever
other fields are important in the earliest Universe], one should be
able to \emph{compute} $\alpha$ from the structure of those fields
and from the detailed geometry of the Weeks manifold. Agreement with
the predictions implied by (\ref{eq:ALPHA}) or (\ref{eq:BETA}) could
be regarded as confirmation of horizon complementarity.

In this work we have proposed that the consequences of cosmic
holography can be understood in a concrete way with the aid of the
Casimir effect. The underlying \emph{reason} for the validity of the
holographic point of view remains to be understood, and this is of
course a major question. It seems possible, for example, that
holography arises here in connection with the peculiarities of
quantum field theory on negatively curved spaces \cite{kn:wilczek}.
Callan and Wilczek observe, for example, that negative curvature
effectively confines even long-range interactions to finite domains,
because of the exponential growth of surface area in hyperbolic
space. This observation, and others in \cite{kn:wilczek}, make the
``effective finiteness" of such spaces seem less surprising; but
much remains to be done to extend such insights to a concrete
understanding of holography and complementarity.

\addtocounter{section}{1}
\section*{\large{\textsf{Acknowledgements}}}
The author is very grateful to Dr Soon Wanmei for help with
numerical work.

\end{document}